# Evaluating the Performance of Existing Full-Reference Quality Metrics on High Dynamic Range (HDR) Video Content

Maryam Azimi, Amin Banitalebi-Dehkordi, Yuanyuan Dong, Mahsa T. Pourazad, and Panos Nasiopoulos

*Abstract*— While there exists a wide variety of Low Dynamic Range (LDR) quality metrics, only a limited number of metrics are designed specifically for the High Dynamic Range (HDR) content. With the introduction of HDR video compression standardization effort by international standardization bodies, the need for an efficient video quality metric for HDR applications has become more pronounced. The objective of this study is to compare the performance of the existing full-reference LDR and HDR video quality metrics on HDR content and identify the most effective one for HDR applications. To this end, a new HDR video dataset is created, which consists of representative indoor and outdoor video sequences with different brightness, motion levels and different representing types of distortions. The quality of each distorted video in this dataset is evaluated both subjectively and objectively. The correlation between the subjective and objective results confirm that VIF quality metric outperforms all to ther tested metrics in the presence of the tested types of distortions[1].

## I. INTRODUCTION

HIGH Dynamic Range (HDR) content has recently received significant recognition in several multimedia application areas. HDR delivers dynamic range that is close to what is perceived by the human visual system (HVS) in real life. HVS is capable of perceiving the light approximately at contrast ratio of $10^5$:1 simultaneously in one scene [1]. This range is far beyond the dynamic range that the majority of existing capturing and display devices are capable of providing. Presently, the vast majority of existing consumer cameras and display devices are able to support Low Dynamic Range (LDR) video content with contrast ratio of approximately 100:1 to 1000:1.

An end-to-end HDR delivery pipeline involves capturing, transmitting, and displaying of HDR content while preserving it brightness and color range. For capturing HDR content that embraces the full visible color gamut and dynamic, one solution is to record the scene with different exposure settings simultaneously and then combine the captured LDR videos to create a single HDR scene [2]. Using this method, the information of both under exposed (dark) and over exposed (bright) areas of the scene is preserved. To display HDR content, one approach is to employ a display system which consists of an LDR LCD panel in front and a LDR projector at the back [3]. Such a system is capable of displaying HDR content with contrast ratio of up to 50000:1 [3].

Given that HDR videos involve much more information than their LDR counterparts, they require a higher number of bits to represent each pixel. Each color component in LDR videos is represented by 8 bits (each pixel has three color components and is represented by 24 bits). Instead, each color component in an HDR video stream is represented in a floating-point notation, which is saved in 10 to 16 bits (depending on the file format) [4-6]. Thus, efficient compression schemes are required for HDR content delivery and storage. Although there is no HDR video compression standard, the existing LDR video compression standards, such as H.264/AVC [7]and HEVC [8-9] may be used to compress the high bit depth color information of HDR content (they can handle up to 12-bit color information). It is worth noting that while these standards may be used for encoding HDR content, they are not optimized for efficiently compressing HDR [10].

The consumer-end quality of the HDR videos depends on how well the quality of data along broadcasting steps (i.e., acquisition, transmission, and display) is preserved. The human observers' opinion of the HDR content quality is the ideal evaluation measure. However, subjective evaluation of multimedia content is not always practical and/or efficient in some applications such as broadcasting and video streaming. In such cases objective quality metrics are required to predict a numerical value for the quality of a video. One approach to evaluate the quality of HDR content is to extend the usage of LDR quality metrics on HDR content. To this end the HDR data needs to be first processed so that its pixel value falls into a range that is supported by LDR quality metrics. This method is known as perceptually uniform (PU) encoding [11]. Another very simple but effective technique for employing LDR metrics on HDR data is based on the multi-exposure inverse tone mapping. In this technique the HDR stream is tonemapped to several LDR streams with different exposure range, and then the LDR metric is applied to each LDR stream and the numerical quality values are averaged at the end [12]. In addition to these LDR quality metric-based approaches, there are a limited number of quality metrics that have been developed specifically for HDR content. Dynamic Range Independent metric (DRI)-VDP [14] and DRI-VQM [15] are two HDR quality metrics that provide a visible difference map. In other words these metrics predict the visibility of the distortions as a map, but they do not generate one single numerical value for quality. However the HDR quality metric proposed by [13] and known as HDR-VDP-2 generates a quality value in addition to the distortion map.

The performance of the most of the above mentioned quality metrics has been tested only on LDR data. One reason might be the lack of a comprehensive HDR video database. To the best of our knowledge, the only existing publicly available HDR video dataset is the one provided by Cad´ık, et al. [16]. The HDR videos in this dataset are designed and rendered for Computer Graphics applications. Their resolution is low (512x512 or lower), they are short (at most 60 frames, 3-second long), and they include scenes with low motion.

The main focus of this paper is to evaluate the performance of the existing LDR and HDR metrics on HDR video content which in turn will provide us with a better understanding of how well each of these metrics work and if they can be applied in capturing, compressing, transmitting process of HDR data. To this end, a comprehensive HDR video database called "DML-HDR" is created and made publicly available to the research community [17]. A series of subjective tests is performed to evaluate the quality of DML-HDR video database when several different representing types of artifacts are present using a HDR display. Then, the correlation between the results from the existing LDR and HDR quality metrics and those from subjective tests is measured to determine the most effective exiting quality metric for HDR.

The rest of this paper is organized as follows: Section II explains the procedure of preparing the HDR video dataset, Section III describes the test setup, Section IV contains the results and discussion, and Section V concludes the paper.

TABLE I
DESCRIPTION OF THE HDR VIDEO DATASET

| Sequence Name | Motion Level | Number of Frames | Environment |
|---|---|---|---|
| Hallway | Intermediate | 253 | Indoor |
| Christmas | Intermediate | 317 | Indoor |
| Playground | Fast | 222 | Outdoor |
| Stranger | Intermediate | 303 | Outdoor |
| Table | Slow | 261 | Indoor |

II. DML-HDR DATASET

One challenge in evaluating HDR video quality is the lack of a representative HDR video dataset. To this end, a comprehensive HDR video database called "DML-HDR" is created [17]. This video dataset consists of five HDR videos all captured by a professional camera capable of capturing HDR videos (RED Scarlet-X), up to 16 bits per each pixel. All videos represent natural scenes. Each video sequence is approximately 10 seconds long with a frame rate of 30 frames per second (fps). All sequences are recorded in 2048×1080 resolution. Table I summarizes the characteristics of each video sequence, and the snapshots of these videos are shown in Fig. 1. Please note that the frames are tone-mapped in Fig. 1 for use on LDR media. This captured videos are available both in RGBE and YUV 12-bit format. RGBE is a lossless HDR video format, where each pixel is encoded with 4 bytes, one byte for red mantissa, one byte for green mantissa, one for the blue mantissa, and one byte for a common exponent [17][19]. The YUV 12-bit format consists of three channels, Y for luma and U and V for Chroma. Each channel is represented by integer values between 0 and 4095 (12 bits).

In the "DML-HDR" video dataset, in addition to the reference video sequences, five distinctive distorted versions of them are also provided for each sequence. The five representative types of distortions applied to each video are listed below:

- *Additive White Gaussian Noise (AWGN):* white Gaussian noise with mean of zero and standard deviation of 0.002 was added to all frames of each video. Based on our knowledge from LDR videos, this value of standard deviation may seem to be too small. However, observations from watching distorted HDR videos on the HDR display showed that AWGN with the standard deviation value of 0.002 is visible. This may be due to their larger dynamic range compared to LDR videos. Note that, before adding the AWGN noise to the content, all pixel values were normalized between 0 and 1. After adding the AWGN noise, pixel values were converted back to the original scale.
- *Mean intensity shift:* the luminance of the HDR videos was globally increased in all the frames of each video sequence by 10% of the maximum scene luminance.
- *Salt and pepper noise:* Salt and pepper noise was added to the 2% of the pixels in each frame of the videos. The distribution of the affected pixels by salt and pepper noise was random.
- *Low Pass Filter*: An 8×8 Gaussian low pass filter with standard deviation of 8 was applied to each frame of all the sequences. Subsequently, rapid changes in intensity in each frame were averaged out.
- *Compression artifacts:* All the videos were encoded using the HEVC encoder (HM software version 12.1) with random access main10 profile configuration. The HEVC encoder settings were as follows: hierarchical B pictures, group of pictures (GOP) size of 8, Internal bit-depth of 12, input video format of YUV 4:2:0 progressive, enabled CABAC entropy coding and rate-distortion optimized quantization (RDOQ). The quantization parameter (QP) was set to 22, 27, 32, and 37 in order to simulate impaired videos with a wide range of compression distortions.

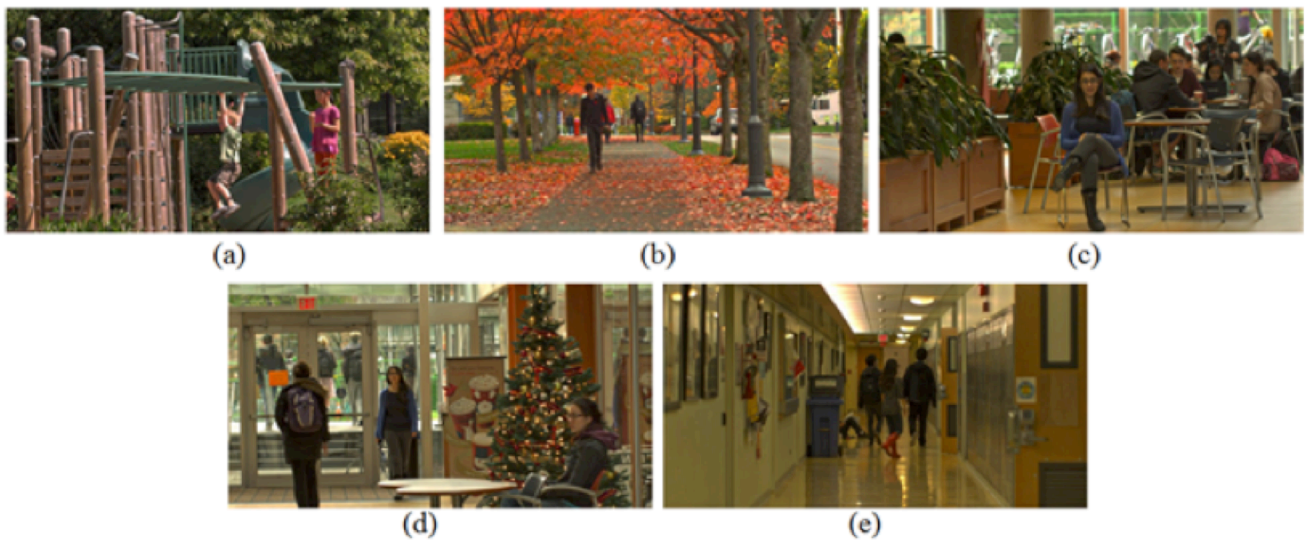

**Fig 1. Snapshots of the first frames of HDR test video sequences (tone-mapped version): (a) Playground, (b) Stranger, (c) Table, (d) Christmas, and (e) Hallway**

The compressed videos are available in 12-bit YUV format in the "DML-HDR" video dataset, whereas all other distorted videos are available in HDR format (.hdr). This is because the YUV format is the default format used by the HEVC reference software (HM software[20]).

## III. EXPERIMENT SETTING

The performance of the objective quality metrics is evaluated by comparing their quality score with the Mean Opinion Score (MOS) on the set of the distorted videos in "DML-HDR" dataset. The following subsections elaborate on the objective and subjective test procedures used in our experiment.

### A. Objective Test Procedure

In order to meaningfully test LDR metrics on HDR content, HDR data has to be adapted in LDR domain. One method for adapting HDR data into the LDR domain is Perceptually Uniform (PU) encoding method [17]. PU encoding method transforms luminance values in the range of $10^5$ cd/m2 to $10^8$ cd/m2 into approximately perceptually uniform LDR values. Another very simple yet effective technique for employing LDR metrics on HDR data is known as multi-exposure method [12]. In this method, the HDR data is tone-mapped with several exposures, uniformly distributed over the dynamic range of the data. The quality of each tone-mapped video, which is in turn an LDR video, is computed by the LDR metric. Then the average of the quality of all the tone-mapped versions forms the actual quality score of the metric. In this test, both methods are applied on the HDR data to be able to test LDR metrics on HDR data. LDR metrics used in our experiment include PSNR, SSIM [21], and VIF [22].

Among the existing HDR metrics, HDR-VDP-2 is used in our experiment, as it is the state-of-the-art full-reference metric that works for all luminance conditions (both LDR and HDR) [13]. This metric is designed based on Daly's visual difference predictor (VDP) [23]. HDR-VDP-2 mimics the human visual system and is designed to predict the visibility of changes caused by artifacts on the test image. The input of the metric includes the reference image, the test image, and some other parameters such as maximum physical luminance of the display, angular resolution of the image, and more options on the viewing environment. The output of the metric is a probability map that determines the probability of detecting dissimilarity between reference and test image by a human observer in each image region. Then by using a pooling strategy the probability map is converted into a value of quality score between 0-100 [13], where 0 represents the lowest quality and 100 stands for the highest quality meaning the reference and test images are identical in terms of the quality. The older versions of HDR-VDP-2 (HDR-VDP 1.7 and HDR-VDP 1.0) provide only a distortion (difference) probability map and do not quantify the visual distortion, thus are not used in our experiment. Similarly, DRI-VDP and DRI-VQM [14] have been excluded from our test, since they only provide a distortion map without a quantitative quality value.

### B. Subjective Test Procedure

The subjective evaluations were conducted in a room complying with the ITU-R BT.500-13 Recommendation [24]. Prior to the actual experiment, a training session was shown to the observers to familiarize them with the rating procedure. The test sessions were designed based on the Double-Stimulus Impairment Scale (DSIS) method [24]. In particular, after each 10-second long reference video, a 3-second gray interval was shown followed by the 10-second long test video. Another 4-second gray interval was allocated after the test video, allowing the viewers to rate the quality of the test video with respect to that of the reference one. The test videos are the distorted videos from "DML-HDR" vide dataset as explained in section II.

The scoring is based on discrete scheme where a numerical value from 1 (worst quality) to 10 (identical quality) is assigned to each test video representing its quality with respect to the reference video [24]. Note that in order to stabilize the subjects' opinion, a few dummy video pairs were presented at the beginning of the test and the subjects were asked to rate them. The collected scores for these videos were discarded from the final results.

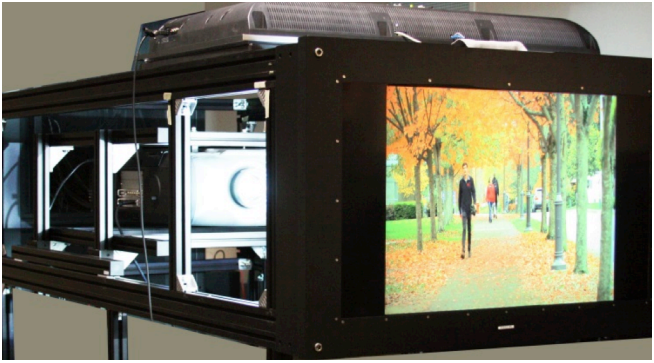

Fig 2. Prototype HDR Display

The videos were displayed on a HDR TV prototype built based on the concept explained in [3]. As illustrated in Fig. 2, this system consists of two main parts: 1) a 40 inch full HD LCD panel in the front, and 2) a projector with HD resolution at the back to provide the backside luminance. The contrast range of the projector is 20000:1. The original HDR video signal is split into two streams, which are sent to the projector and the LCD (see [3] for details). The input signal to the projector includes only the luminance information of the HDR content and the input signal to the LCD includes both luma and chroma information of the HDR video. Using this configuration, the light output of each pixel is effectively the result of two modulations with the two individual dynamic ranges multiplied, yielding an HDR signal. This HDR display system is capable of emitting light at a maximum brightness level of 2700 $cd/m^2$.

Eighteen adult subjects including 10 males and 8 females participated in our experiment. The subjects' age range was from 19 to 35 years old. Prior to the tests, all the subjects were screened for color blindness using the Ishihara chart and visual acuity using the Snellen charts. Those subjects that failed the pre-screening test did not participate in the test.

## IV. RESULTS AND DISCUSSIONS

After collecting the subjective results, the outlier subjects were detected according to the ITU-R BT.500-13 recommendation in [24]. No outlier was detected in this test. The Mean Opinion Score (*MOS*) for each impaired video was calculated by averaging the scores over all the subjects with 95% confidence interval.

Table II summarizes the results of the correlation between the objective quality scores and the ones of the subjective tests. In order to estimate each metric's accuracy, the Pearson Linear Correlation Coefficient (PCC) is calculated between MOS values and the obtained objective quality indices. The Spearman Rank Order Correlation Coefficient (SCC) is also computed to estimate the monotonicity in the metrics' results. The PCC and SCC in each column are calculated over the entire video data set. The results are reported based on three impairments categories: a) compression artifacts, b) AWGN, intensity shifting, salt & pepper noise, and low pass filtering, and c) all the impairments used in our study.

As it is observed from Table II, in the presence of the AWGN, intensity shifting, salt & pepper noise, and low pass filtering distortions, VIF yields the best performance compared to HDR-VDP-2 and other used LDR metrics in our experiment (regardless of the employed adaptation method, i.e., Multi-exposure or PU encoding). In the presence of the compression artifacts, however, HDR-VDP-2 outperforms all other tested metrics. Overall, in the presence of the tested distortions, VIF with PU encoding shows the best performance in predicting the quality of the HDR videos compared to other tested metrics.

## V. CONCLUSION

The main purpose of the paper was to investigate the performance of existing quality metrics in evaluating the quality of HDR content. To this end a representative HDR dataset is captured and several types of impairments are applied. The dataset included 40 test videos with five types of distortions. Subject standardized subjective test procedure is implemented. In the experiment, not only HDR quality metrics, but also the proposed schemes based on LDR quality metrics are used to predict the quality of HDR videos. Experiments results showed that in the presence of compression distortions, HDR-VDP2 outperforms all other metrics. Overall VIF using PU encoding yields the best performance in the presence of all the tested impairments.

TABLE II
**CORRELATION OF SUBJECTIVE RESPONSES WITH PREDICTION OF OBJECTIVE QUALITY METRICS**

| Metric/Method | Impairments: AWGN, Intensity shifting, salt & pepper noise, and low pass filtering | | Impairment: Different levels of compression, QP: 22, 27, 32, 37 | | Impairment: All | |
|---|---|---|---|---|---|---|
| | *Pearson Correlation* | *Spearman Correlation* | *Pearson Correlation* | *Spearman Correlation* | *Pearson Correlation* | *Spearman Correlation* |
| HDR-VDP-2 | 0.3639 | 0.3686 | **0.9270** | **0.8113** | 0.4871 | 0.3413 |
| PSNR (PU encoding) | 0.6754 | 0.4122 | 0.7444 | 0.7355 | 0.6361 | 0.7096 |
| SSIM (PU encoding) | 0.5634 | 0.5004 | 0.8881 | 0.7470 | 0.4526 | 0.5146 |
| VIF (PU encoding) | **0.9723** | **0.8703** | 0.8490 | 0.7929 | **0.8522** | **0.8462** |
| PSNR (Multi-Exposure) | 0.8631 | 0.4799 | 0.7744 | 0.6163 | 0.5180 | 0.7303 |
| SSIM (Multi-Exposure) | 0.7065 | 0.4724 | 0.8932 | 0.6988 | 0.5400 | 0.5198 |
| VIF (Multi-Exposure) | **0.92981** | **0.7273** | 0.7842 | 0.6830 | 0.6450 | 0.7517 |